\documentclass[aps,prb,twocolumn,showpacs]{revtex4}
\usepackage{graphicx}
\usepackage{latexsym}
\usepackage{amsmath}
\usepackage{subfigure}
\usepackage{natbib}
\usepackage{inputenc}
\usepackage[usenames,dvipsnames]{color}

\begin{document}
\title{Wannier interpolation of the electron-phonon matrix elements in
  polar semiconductors: Polar-optical coupling in GaAs}
\author{J. Sjakste}
\email{jelena.sjakste@polytechnique.edu}
\author{N. Vast}
\affiliation{Ecole Polytechnique, Laboratoire des Solides Irradi\'es,
  CEA-DSM-IRAMIS, CNRS UMR 7642, Paris-Saclay University, 91120 Palaiseau,
  France}
\author{M. Calandra}
\author{F. Mauri}
\affiliation{IMPMC, UMR CNRS 7590, Sorbonne Universit\'es - UPMC Univ. Paris 06, MNHN, IRD, 4 Place Jussieu,
F-75005 Paris, France}

\begin{abstract}
We generalize the Wannier interpolation of the electron-phonon matrix
elements to the case of  polar-optical coupling in polar semiconductors. 
We verify our methodological developments against experiments,
by  calculating the widths of the electronic bands due to
electron-phonon scattering in GaAs,
the prototype  polar semiconductor. 
The calculated widths are then used to estimate the broadenings of
excitons at critical points in GaAs
and the electron-phonon relaxation times of hot electrons. Our
findings are in  good agreement with available experimental data. 
Finally, we demonstrate that while the Fr\"ohlich
interaction is the dominant scattering process for electrons/holes close to the valley minima, in agreement with low-field 
transport results, at higher energies, the intervalley scattering dominates the relaxation dynamics of hot electrons 
or holes. The capability of interpolating the polar-optical coupling
opens new perspectives in the calculation of
optical absorption and transport properties in semiconductors and thermoelectrics.

\end{abstract}
\maketitle

\date{\today}

\maketitle

\section{Introduction}
\label{sec:intro}
Electron-phonon coupling plays a fundamental role in the relaxation of
photoexcited electrons, thus affecting the performance 
of photovoltaic~\cite{Polman:2012} and other semiconductor-based devices~\cite{Delerue:2004,Zebarjadi:2012}.
In many cases, the electron-phonon coupling determines the
magnitude of the lifetimes of electronic states inside the 
band gap, and the widths of the corresponding absorption peaks~\cite{Zollner:1989:GaP,Li:1994,Steger:2009}.
It is also responsible for shifts and broadenings of the interband 
critical points in semiconductors~\cite{Gopalan:1987,Lautenschlager:1987}. 
The interpretation and analysis of the (ultra)fast dynamics of relaxation of photoexcited electrons is particularly difficult, because many relaxation
processes are present simultaneously, and disentangling their
respective contributions requires 
\textit{ad hoc} information on their relative importance and order of magnitude \cite{Rossi:2002}.

At the same time, \textit{ab initio} calculations provide an effective tool
to estimate electron-phonon coupling strength in metals~\cite{Mauri:1996:a,Giustino:2007,
Calandra:2011} and semimetals like graphene\cite{Piscanec:2004,Lazzeri:2006,Calandra:2007,Bonini:2007} or bismuth~\cite{Papalazarou:2012,Faure:2013}. 
In the case of semiconductors, the predictive capability of calculations based 
on density functional perturbation theory (DFPT)~\cite{Baroni:1987,Baroni:2001}
 for the electron-phonon
matrix elements has been demonstrated in a number of semiconductors~\cite{Sjakste:2007c,Tyuterev:2011,Sjakste:2013b}, 
alloys~\cite{Murphy-Armando:2008,Vaughan:2012} and nanostructures~\cite{Murphy-Armando:2011b,Sjakste:2013b}. 

Recently, a method to interpolate the electron-phonon coupling matrix
elements using Wannier fuctions
has been introduced~\cite{Giustino:2007,Calandra:2010,Marzari:2012},
providing a computationally efficient method to calculate electron-phonon matrix elements
on extremely fine grids in the Brillouin zone (BZ) of metals. This has proved to be crucial to predict various material properties
such as for example nonadiabaticity \cite{Calandra:2010} or superconductivity \cite{Casula:2011,Margine:2013}. 

The method~\cite{Giustino:2007,Calandra:2010}  has also been used to increase the precision of integrals 
related to electron-phonon scattering times and scattering rates in semiconductors~\cite{Bernardi:2014}, 
being, however, limited to nonpolar semiconductors. 
Indeed, in polar semiconductors, a long-wavelength longitudinal optical (LO) phonon induces an electric field, and 
the interaction of electrons with this macroscopic electric field -~the polar-optical coupling  or  Fr\"ohlich interaction~- is divergent when the
phonon wave vector tends to zero. As the electron-phonon matrix
elements  related to the Fr\"ohlich interaction are not localized in
the  Wannier basis, these matrix elements cannot be properly interpolated with the method
of Refs.~\onlinecite{Giustino:2007,Calandra:2010}.

In this work, we extend the method~\cite{Giustino:2007,Calandra:2010} to take into account polar-optical coupling, 
and apply it to GaAs which is an archetype of a polar semiconductor.
First, we present  the theoretical background of  our method, which we
validate by comparing the Wannier interpolated electron-phonon matrix elements
with that obtained by direct calculation within DFPT. Next, we calculate  
band broadenings due to the electron-phonon interaction 
for the highest valence and lowest conduction states. The 
calculated broadenings represent the total probability of the momentum relaxation
of the hot electrons or holes due to electron-phonon interaction,
in good agreement with recent pump-probe experiments \cite{Kanasaki:2014}.
We analyze the role of the Fr\"ohlich interaction in the relaxation
of excited electrons in GaAs, and we find that the Fr\"ohlich interaction is responsible of the
quasi-totality of the electron-phonon relaxation rates at low excitation energies, while representing only 
 10\% of the electron-phonon relaxation rates for hot electrons or holes.   
The calculated data have then been used to estimate the
broadenings of the $E_1$ and $E_2$ critical points (CP) in GaAs. 
For $E_1$, the calculated broadening is in  very good agreement 
with the experimental results of work~\onlinecite{Gopalan:1987} and with previous calculations\cite{Gopalan:1987}.
 Finally, for $E_2$, the calculated broadening  are in satisfactory agreement with experimental results,  in contrast with 
the previous calculation with the empirical pseudopotential method~\cite{Gopalan:1987}. 
 
\section{Theory}
\label{sec:theor}

\subsection{Electron-phonon matrix element}

The matrix element $\mathbf{d}_{mn}^s(\mathbf{k},\mathbf{k}+\mathbf{q})$  
of the 
periodic part of the static and self-consistent response potential 
 for a monochromatic perturbation of $\mathbf{q}$ wave vector,  $\mathbf{u}_{\mathbf{q}s}$, reads: 
\begin{equation}
\mathbf{d}_{mn}^s(\mathbf{k},\mathbf{k}+\mathbf{q})= \langle \mathbf{k} n 
|\frac{\delta v_{SCF}}{\delta \mathbf{u}_{\mathbf{q}s}}|\mathbf{k}+\mathbf{q} m \rangle, 
\label{d_element}
\end{equation}
where  $|\mathbf{k} n\rangle$ stands for the periodic  part of the Bloch wave function of the initial electronic state, \textit{i.e.}
$|\psi_{\mathbf{k},n}\rangle=e^{i\mathbf{k}\mathbf{r}}|\mathbf{k} n\rangle/\sqrt{N_k}$. 
The vector  $\mathbf{k}$ is the electronic wave vector,
 $\mathbf{q}$ is the phonon wave vector, and
$n$ and $m$ are the band numbers of the initial and final states. $N_k$ is the number of points in the $\mathbf{k}$-grid on which
$\psi_{\mathbf{k},n}$ are generated, the periodic part of the wave-function being normalized in  
the unit cell.
$\mathbf{u}_{\mathbf{q}s}$ is the Fourier transform of the phonon
displacement of atom $s$.
The quantity $\delta v_{SCF} /\delta  \mathbf{u}_{\mathbf{q}s}$ is the periodic part of the (static and self-consistent) response potential.

The electron-phonon matrix element $g_{nm}^{\nu}$ reads:
\begin{equation}
 g_{nm}^{\nu}(\mathbf{k},\mathbf{k}+\mathbf{q}) = \sum_s \mathbf{e}^{s\nu}(\mathbf{q})
\mathbf{d}_{mn}^s(\mathbf{k},\mathbf{k}+\mathbf{q})/\sqrt{2M_s\omega_{\mathbf{q}\nu}}.
\label{g_element}
\end{equation}

We have used $\mathbf{e}^{s\nu}(\mathbf{q})$ for the phonon
eigenvector ($s$ labels the atoms in the unit cell, $\nu$ labels the phonon mode), $\omega_{\mathbf{q}\nu}$ is the phonon
frequency, and $M_s$ is the atomic mass.

In analogy with our previous works~\onlinecite{Sjakste:2013,Sjakste:2007} and the original work~\onlinecite{Zollner:1990}, the 
\textit{deformation potential} for an individual
transition is defined
as a quantity proportional to the absolute value of the electron-phonon matrix element of eq.~(\ref{g_element}):
\begin{equation}
D_{nm}^{\nu}(\mathbf{k},\mathbf{k}+\mathbf{q})=\frac{\sqrt{2\rho\Omega \omega_{\mathbf{q}\nu} }}{\hbar}|g_{nm}^{\nu}(\mathbf{k},\mathbf{k}+\mathbf{q})|
.
\label{DefPot}
\end{equation}
Here, $\rho$ is the mass density of the crystal, and $\Omega$ is the crystal volume. In the case of several initial and/or final
electronic bands, we define the  total deformation potential as:
\begin{equation}
D_{tot}^{\nu}=\sqrt{\Sigma_{nm}(D_{nm}^{\nu})^2}. 
\label{Total_DefPot}
\end{equation}

The Wannier interpolation of electron-phonon matrix elements was first
introduced in Ref.~\onlinecite{Giustino:2007}. Implementation of the Wannier interpolation procedure into the \textsc{Quantum ESPRESSO} 
package~\cite{Baroni:2001}, which we have used in this work, was described in Ref.~\onlinecite{Calandra:2010}.
In this work, we repeat only the part of the comprehensive description of Ref.~\onlinecite{Calandra:2010} that  is 
necessary for the understanding of the extension to polar-optical coupling introduced in the next subsection.

A set of Wannier functions centered on site $\mathbf{R}$ are defined by the relation:
\begin{equation}
|\mathbf{R}m\rangle=\frac{1}{\sqrt{N_k}}\sum_{\mathbf{k}n}e^{-i\mathbf{k}\mathbf{R}}U_{nm}(\mathbf{k})|\psi_{\mathbf{k}n}\rangle
.
\end{equation}
A transformation matrix, $U_{mn}(\mathbf{k})$, is determined by the Wannierization procedure (see Ref.~\onlinecite{Calandra:2010}).

The matrix elements $\mathbf{d}_{mn}^s(\mathbf{k},\mathbf{k}+\mathbf{q})$ are calculated within
DFPT. As emphasized in Ref.~\onlinecite{Calandra:2010}, the periodic parts $|\mathbf{k}n\rangle$
and $|\mathbf{k}+\mathbf{q}m\rangle$ have to be exactly the same wavefunctions used for the Wannierization
procedure. This allows to fix their arbitrary phases appearing in the $|\mathbf{k}n\rangle$'s because of
the numerical routine used for the diagonalization of matrixes containing complex numbers, or other numerical reasons \cite{Calandra:2010}. 

The matrix element $\mathbf{d}$ in the Wannier function basis is obtained by Fourier transform as:
\begin{eqnarray}
\mathbf{d}_{mn}^s(\mathbf{R},\mathbf{R}_L)=\frac{1}{N_k}\sum_{\mathbf{k},\mathbf{q}}^{N_k}\sum_{m',n'}
e^{-i\mathbf{k}\mathbf{R}+i\mathbf{q}\mathbf{R_L}}\,
\tilde{\mathbf{d}}_{m'n'}^s(\mathbf{k}+\mathbf{q},\mathbf{k}) \nonumber\\
\label{elphon_in_wan}
\end{eqnarray}
with
\begin{eqnarray}
\tilde{\mathbf{d}}_{m'n'}^s(\mathbf{k}+\mathbf{q},\mathbf{k}) =U_{mm'}^*(\mathbf{k}+\mathbf{q})\mathbf{d}_{m'n'}^s(\mathbf{k}+\mathbf{q},\mathbf{k})
U_{n'n}(\mathbf{k})\nonumber\\
\end{eqnarray}

Finally, when the localization conditions are verified on $\mathbf{d}_{mn}^s(\mathbf{R},\mathbf{R}_L)$ (see Ref.~\onlinecite{Calandra:2010}),
one can obtain, by a slow Fourier transform, $\mathbf{d}_{mn}^s(\mathbf{k}+\mathbf{q},\mathbf{k})$ with $\mathbf{k}$ and $\mathbf{q}$ being
any points in Brillouin zone:
\begin{eqnarray}
\mathbf{d}_{mn}^s(\mathbf{k}+\mathbf{q},\mathbf{k})
=\frac{1}{(N_k)^2}\sum_L\sum_{\mathbf{R}}
\sum_{m'n'}e^{i\mathbf{k}\mathbf{R}+i\mathbf{q}\mathbf{R}_L}\nonumber\\
U_{m'm}(\mathbf{k}+\mathbf{q})\mathbf{d}_{m'n'}^s(\mathbf{R},\mathbf{R}_L)
U_{nn'}^*(\mathbf{k})
\label{elphon_back}
\end{eqnarray}

In this work, GaAs is described within the local density approximation (LDA), and with the
same pseudopotentials as in our previous works~\onlinecite{Botti:2002,Sjakste:2007}. For the electronic density calculation, we used an energy cutoff value 
of 45 Ry and a Monkhorst-Pack grid of  $12\times12\times12$ points in the BZ. The Wannier interpolation of the structure was carried out using
$6\times6\times6$ and $8\times8\times8$ $\mathbf{k}$-point grids centered at $\Gamma$, with ten Wannier functions and 45 DFT Bloch wavefunctions. The large number 
of Wannier functions and DFT bands, as well as the rather dense $\mathbf{k}$-point grids, are related to the costly disentanglement procedure necessary 
to satisfactorily reproduce the lowest conduction bands of GaAs.

The same $6\times6\times6$ and $8\times8\times8$ grids centered at $\Gamma$ were used as initial grids to calculate  electron-phonon matrix elements
within DFPT, which were then Wannier-interpolated using the interpolation method extended to polar-optical coupling described in the next paragraph.

\subsection{The long-range Fr\"{o}hlich interaction}

\begin{figure}
\begin{center}
\includegraphics[width=7.cm]{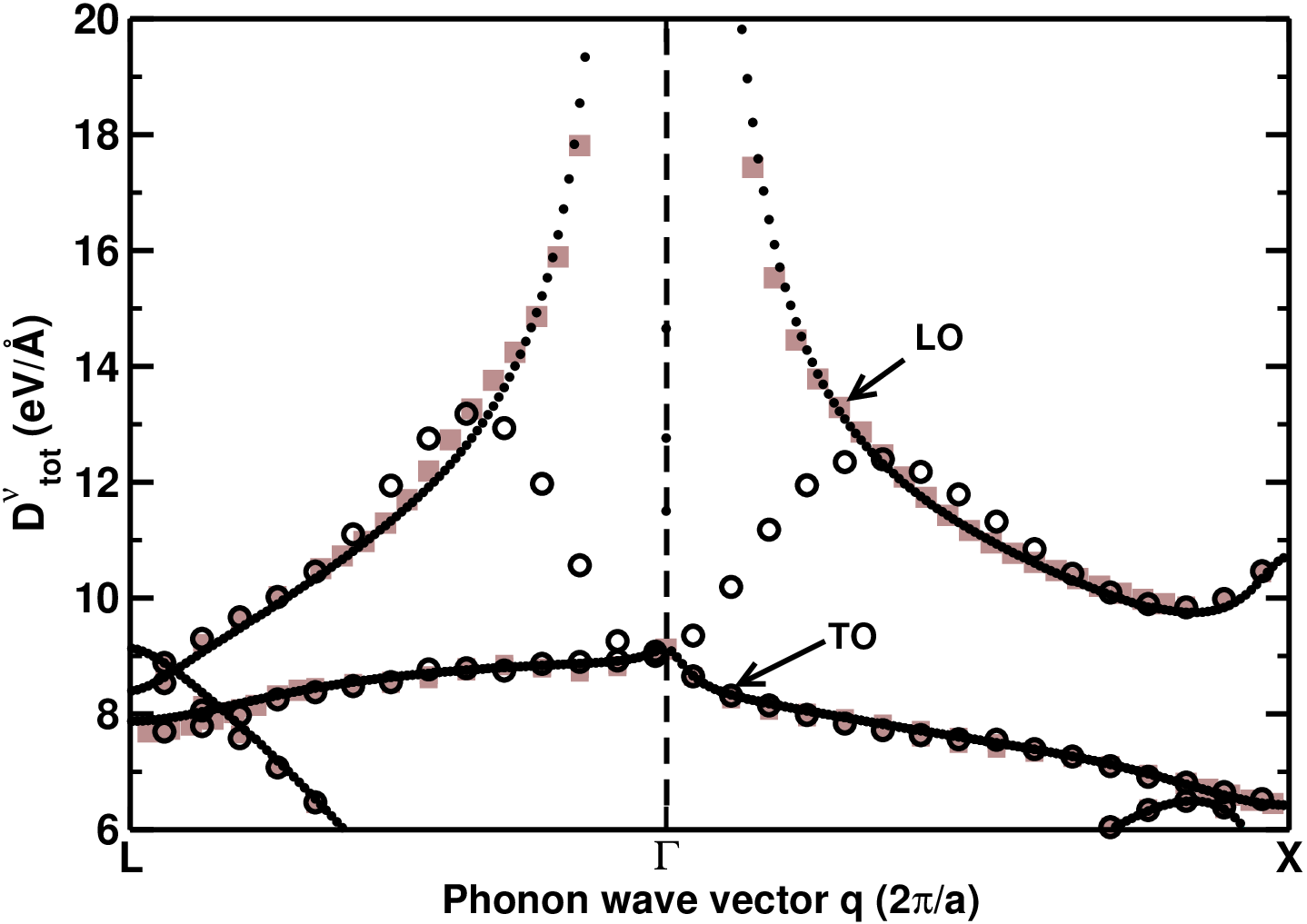}
\caption{GaAs. Deformation potentials for the total contribution of the three highest valence bands, see Eq.~(\ref{Total_DefPot}). 
The initial electronic state is at $\mathbf{k}=\Gamma$, and the phonon wave vector $\mathbf{q}$ is varying 
along high symmetry lines in the BZ. Black line: direct DFPT calculation. Grey squares: Wannier-interpolation extended to polar-optical coupling.
Circles: standard Wannier interpolation for metals and non-polar
semiconductors.}
\label{fig:TO}
\end{center}
\end{figure}

The Fr\"ohlich interaction is long-range, and thus the electron-phonon matrix elements
for long-wavelength longitudinal optical phonons in polar materials
are not localized in the real-space Wannier basis.
A proof is given in Fig.~\ref{fig:TO}, where the interpolation method of Refs.~\onlinecite{Giustino:2007,Calandra:2010} is shown to 
fail for LO phonon as $\mathbf{q}\rightarrow 0$. Indeed, as one can see from Fig.~\ref{fig:TO}, at large  $|\mathbf{q}|$ vectors,
the character of the  electron-phonon matrix elements is completely short-range for all phonon branches including the LO branch
and is well reproduced by the
standard interpolation method.~\cite{Calandra:2010}. At small  $|\mathbf{q}|$ vectors however,
instead of the characteristic $1/|\mathbf{q}|$ behaviour, the interpolation method of Ref.~\onlinecite{Calandra:2010}
yields the same values of deformation potentials for the LO and TO branches.
This is because, for the highest valence bands,  the electron-phonon interaction with long-wavelength
LO phonons at $\mathbf{k}=\Gamma$ contains both long-range (Fr\"ohlich) and short-range contributions~\cite{Yu:Cardona:2001}.
At vanishing $|\mathbf{q}|$, the short-range contribution is the same for the LO and TO phonons.

The absence of LO/TO splitting within a standard Wannier interpolation procedure is in analogy with the LO/TO splitting of phonon frequencies in polar semiconductors.
Indeed, if dynamical matrices are interpolated with a real space cutoff~\cite{Giannozzi:1991,Baroni:2001}, 
the LO/TO splitting of the phonon modes is absent, and the long-range part of the dynamical matrix needs to be subtracted before
Fourier-interpolation into the real space and re-introduced afterwards in order to reproduce the  LO/TO splitting~\cite{Giannozzi:1991,Baroni:2001}. 
We apply similar scheme in the case of electron-phonon matrix elements, with 
the nonlocal part of the electron-phonon interaction represented  by the Vogl model described in the next paragraph.

\subsection{The Vogl model}

\begin{figure}
\begin{center}
\includegraphics[width=7.cm]{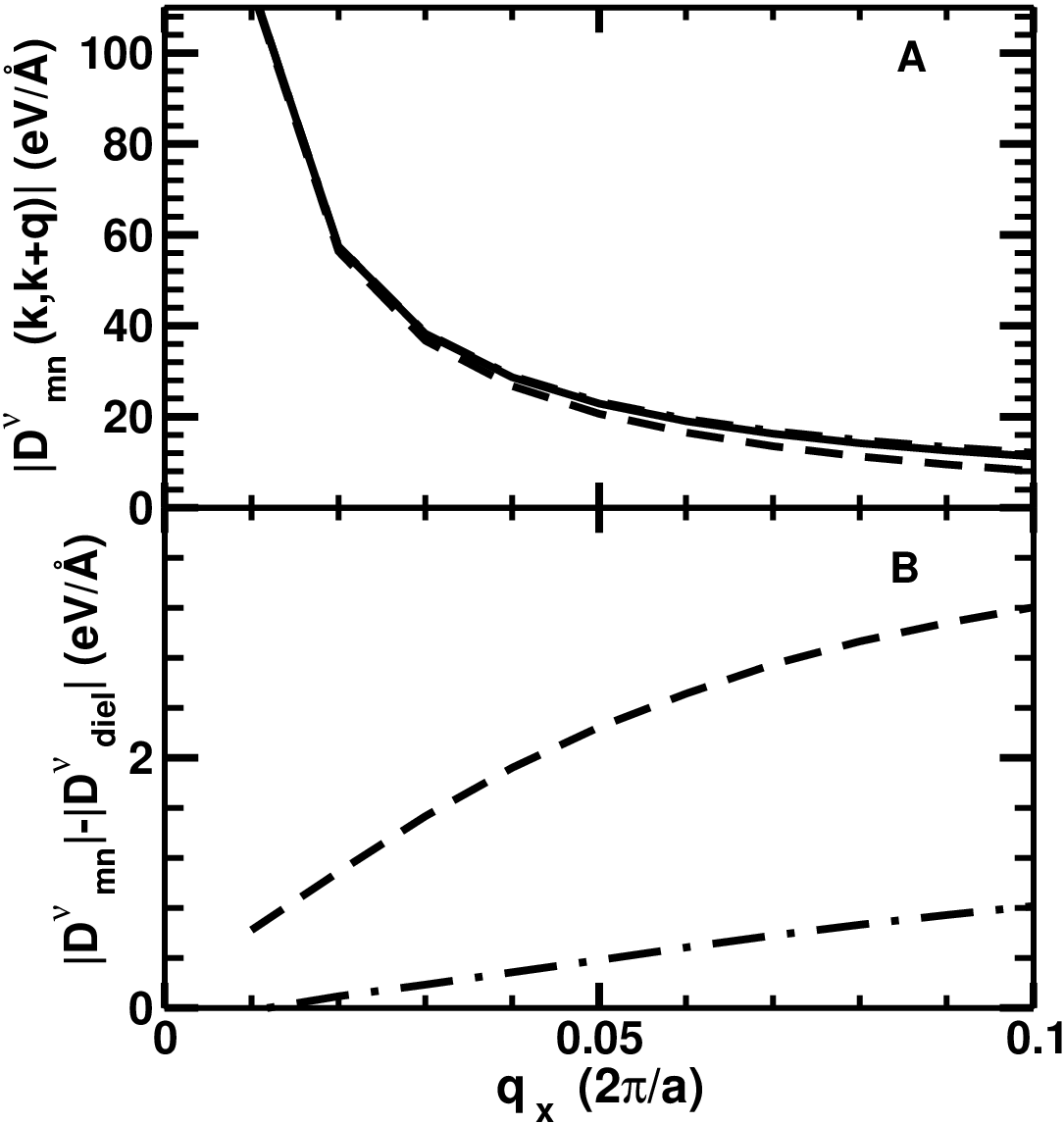}
\caption{GaAs. LO deformation potentials for the initial electronic state with
$\mathbf{k}=X$ and the phonon wave vector changing along the (100) direction: $\mathbf{q}=(q_x,0,0)2\pi/a$, for $q_x \ge$  0.01.
Panel A (top): deformation potentials calculated within DFPT for the lowest conduction band (dashed line) and two lowest valence bands (dot-dashed line, visible in panel B), compared with the deformation potentials calculated from the model of eq.~(\ref{diel}) (solid line). 
Panel B (bottom): The difference between the deformation potentials
calculated with DFPT, and the model of eq.~(\ref{diel}). }
\label{fig:model}
\end{center}
\end{figure}

The electron-phonon matrix element of the
interaction of electrons with the macroscopic electric field induced by long-wavelength longitudinal optical phonons
(the Fr\"ohlich interaction) was derived by Vogl in Ref.~\onlinecite{Vogl:1976}, for long-wavelength
phonons ($\mathbf{q}\rightarrow 0$). The leading contribution, in terms of ascending powers of $q$, to the
intraband electron-phonon matrix element, is given by the interaction with  a dipole potential ($\propto \frac{1}{|q|}$) screened by the 
high-frequency 
dielectric
tensor $\mathbf{\epsilon}_{\infty}$ (see Eq. 3.12 of Ref.~\onlinecite{Vogl:1976}), and reads:
\begin{equation}
g_{diel}^{\nu}(\mathbf{q})=\frac{4\pi i e}{\mathbf{q}\cdot\overleftrightarrow{\mathbf{\epsilon}}_\infty
\cdot\mathbf{q}}\sum_{s}\sum_{\lambda'} q_{\lambda^{'}} Z_{\lambda^{\prime}\lambda s}\mathbf{e}^{s\nu}_{\lambda}(\mathbf{q})/\sqrt{2M_s\omega_{\mathbf{q}\nu}}
\label{diel}
\end{equation}
The corresponding deformation potential then becomes:
\begin{equation}
D^{\nu}_{diel}(\mathbf{q})=\frac{\sqrt{2\rho\Omega\omega_{\mathbf{q}\nu}}}{\hbar}|g_{diel}^{\nu}(\mathbf{q})|
\label{diel-defpot}
\end{equation}
In eq. (\ref{diel}), $e$ is electronic charge, 
$Z_{\lambda^{'}\lambda s}$ is the Born effective charges tensor for
atom $s$, and $\lambda, \lambda'$ denotes the cartesian components.  Among other
approximations, it was assumed, in Ref.~\onlinecite{Vogl:1976}, that:
\begin{equation}
\langle \mathbf{k}+\mathbf{q}m|\mathbf{k}n\rangle = \delta_{mn}+O(q^2).
\label{approx}
\end{equation}
Note that such a relation assumes a smooth and analytic relative phase relation among the
$|\mathbf{k},n\rangle$ and $|\mathbf{k+q},m\rangle$ states. Such a requirement is satisfied by the
phase choice giving the localised Wannier functions, but not by the arbitrary phase given by the
diagonalisation procedure.
Furthermore, eq. (\ref{approx})
is the reason why the expression~(\ref{diel}) does not depend on the
electronic wave vector $\mathbf{k}$ nor on the band indexes of the
initial and final electronic states (which are assumed to be the
same).  Nevertheless, expression~(\ref{diel}) describes well the
asymptotic behaviour of the electron-phonon matrix elements in polar
semiconductors as $\mathbf{q}\rightarrow 0$, as one can see in
Fig.~\ref{fig:model}, where the behaviour of the electron-phonon
matrix elements for the lowest conduction band (dashed line) and for
the highest valence bands (dot-dashed line) of GaAs calculated within
DFPT are compared with the equations~(\ref{diel}) and (\ref{diel-defpot}) of the model for the
Fr\"ohlich interaction along the (100) direction in the Brillouin zone
(solid line).  At large $q_x$, the short-range character of the
electron-phonon matrix elements is different for the conduction band
and the valence bands, and cannot be described with the model of
Eq.~(\ref{diel}). At small $q_x$, on the contrary, the asymptotic
behaviour of the electron-phonon matrix elements for the valence bands
and the conduction band becomes similar, and this behaviour is
extremely well described by the model of eq.~(\ref{diel}).

\subsection{Wannier interpolation extended to polar-optical coupling}

The method we propose in order to extend the interpolation method of the electron-phonon matrix elements
to polar semiconductors is similar to the one described in Ref.~\onlinecite{Giannozzi:1991} for the interpolation
of the force constants in polar materials. 
The idea is that the long-range contribution to the electron-phonon matrix elements, described with
the model of eq.~\ref{diel},  has to 
be subtracted from the electron-phonon matrix elements before the Fourier transform to real space in eq.~(\ref{elphon_in_wan})
and restored after the Fourier transform back to the reciprocal space in eq.~(\ref{elphon_back}).

We use the Ewald sum in order to take into account the periodicity properties
of the crystal~\cite{Giannozzi:1991} and define:
\begin{eqnarray}
d^{diel}_{\lambda s}(\mathbf{q})
=\nonumber\\
4\pi i e\sum_{\lambda^{'}}\sum_{\mathbf{G}}
\frac{e^{-(\mathbf{q}+\mathbf{G})^2/4\alpha}}{(\mathbf{q}+\mathbf{G})\cdot\overleftrightarrow{\mathbf{\epsilon}}_\infty\cdot(\mathbf{q}+\mathbf{G})}
Z_{\lambda^{\prime}\lambda
s}(q_{\lambda^{'}}+G_{\lambda^{'}}) \nonumber\\
\label{diel_matr}
\end{eqnarray}
Here, $\alpha$ 
is a convergence parameter (we used $\alpha=$ 5 $(\frac{2\pi}{a})^2$ in this work ), 
and the $\mathbf{G}$ are the reciprocal lattice vectors. The dielectric term, eq.~(\ref{diel_matr}), depends only on the
phonon wave vector $\mathbf{q}$, and on material characteristics such
as the Born effective charges $Z_{\lambda^{'}\lambda s}$ and the dielectric constant $\mathbf{\epsilon}_{\infty}$, 
which are calculated within linear response theory~\cite{Baroni:2001}.

Then we subtract $ d^{diel}_{\lambda}(\mathbf{q})$ from the intraband
Fourier transform of the deformation potential in the optimally smooth subspace,
namely we define:
\begin{eqnarray}
{\tilde {\cal D}}_{m'n'}^{\lambda
  s}(\mathbf{k}+\mathbf{q},\mathbf{k})=\tilde{d}_{m'n'}^{\lambda s}(\mathbf{k}+\mathbf{q},\mathbf{k})-\delta_{m',n'}
d^{diel}_{\lambda s}(\mathbf{q})
\nonumber\\
\end{eqnarray}

We then carry out the transformation in Eq. \ref{elphon_back} on the
matrix  ${\bf {\tilde {\cal D}}}_{m'n'}^s(\mathbf{k}+\mathbf{q},\mathbf{k})$,
\begin{eqnarray}
\mathbf{{\cal D}}_{mn}^s(\mathbf{k}+\mathbf{q},\mathbf{k})=
\frac{1}{(N_k)^2}\sum_L\sum_{\mathbf{R}}
\sum_{m'n'}e^{i\mathbf{k}\mathbf{R}+i\mathbf{q}\mathbf{R}_L}\nonumber\\
U_{m'm}(\mathbf{k}+\mathbf{q}) {\bf {\tilde{\cal D}}}_{m'n'}^s(\mathbf{R},\mathbf{R}_L)
U_{nn'}^*(\mathbf{k})
\label{elphon_back_D}
\end{eqnarray}
where ${\bf {\tilde{\cal D}}}_{m'n'}^s(\mathbf{R},\mathbf{R}_L)$ is
obtained via Eq. \ref{elphon_in_wan} with
$\tilde{\mathbf{d}}_{m'n'}^s(\mathbf{k}+\mathbf{q},\mathbf{k})$
replaced by
${\tilde {\cal D}}_{m'n'}^s(\mathbf{k}+\mathbf{q},\mathbf{k})$.

Finally we add back $d^{diel}_{\lambda s}(\mathbf{q})$ where now ${\bf q}$
is any phonon wave vector in the Brillouin zone, namely
\begin{eqnarray}
d_{mn}^{\lambda s}(\mathbf{k}+\mathbf{q},\mathbf{k})=
{\cal D}_{mn}^{\lambda s}(\mathbf{k}+\mathbf{q},\mathbf{k})+
d^{diel}_{\lambda s}(\mathbf{q})
\delta_{m,n} 
\end{eqnarray}

In figure~\ref{fig:TO},  the results obtained with the  method of Wannier interpolation extended to polar-optical coupling are shown in grey squares. 
The behaviour of deformation potentials corresponding to LO phonons is well reproduced by the Wannier interpolation extended to polar-optical coupling, 
in contrast with the standard Wannier interpolation method.

\begin{figure}
\begin{center}
\includegraphics[width=7.5cm]{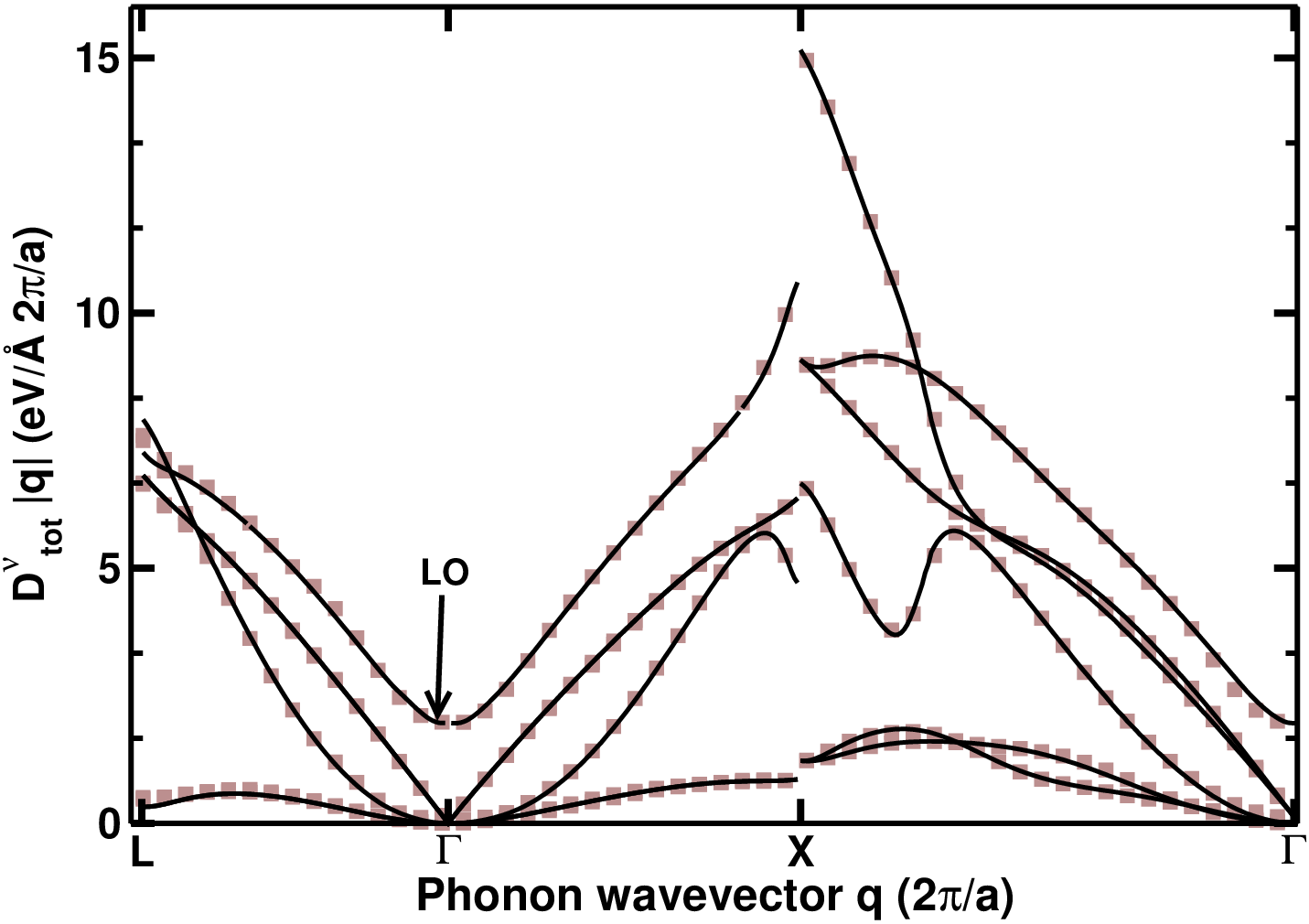}
\caption{
GaAs. Deformation potentials for the total contribution of the three highest valence bands, see eq.~(\ref{Total_DefPot}), 
multiplied by the modulus of the phonon wave vector $|\mathbf{q}|$. 
The initial electronic state is at $\mathbf{k}=\Gamma$, and the phonon wave vector $\mathbf{q}$ is varying
along high symmetry lines in the BZ. Black line: direct DFPT
calculation. Grey squares: our results obtained with the Wannier
interpolation extended to the polar-optical coupling.
}
\label{fig:matel}
\end{center}
\end{figure}

In figure~\ref{fig:matel}, we show the total deformation potentials for the three highest valence bands of GaAs (${n,m}=2,3,4$),
multiplied by the modulus of the phonon wave vector $|\mathbf{q}|$, for all six phonon modes of GaAs. The crystal momentum of the initial
electronic state was taken to be  $\mathbf{k}=\Gamma$, while  the wave vector of the final electronic state $\mathbf{k+q}$ changes as the 
phonon wavector $\mathbf{q}$ varies along high symmetry lines in the BZ. We chose to multiply  deformation potentials
by the  modulus of $\mathbf{q}$, as, due to the Fr\"ohlich interaction,
the deformation potential for the LO phonon tends to infinity as $\frac{1}{|q|}$ and thus the values are very high close to $\Gamma$. In black are represented reference DFPT values, and
in grey squares are shown the deformation potentials which were interpolated using our Wannier interpolation method extended to  polar-optical coupling. As one can see, the
agreement between DFPT calculations and the Wannier-interpolated  deformation potentials is excellent. The non-zero value
of the  deformation potential multiplied by $\mathbf{q}$ at $\mathbf{q}=\Gamma$ is due to the diverging LO-phonon Fr\"ohlich interaction,
which is now properly described. 

In conclusion, the method of Wannier interpolation extended to polar-optical coupling yields interpolated electron-phonon matrix 
elements with the same precision as the "standard" one at large phonon $\mathbf{q}$ vectors, but, in contrast to
the standard method, it enables us to correctly describe the diverging LO-phonon Fr\"ohlich interaction at vanishing $\mathbf{q}$.

\section{Results }

\subsection{Scattering rates and role of the Fr\"ohlich interaction}

\begin{figure}
\begin{center}
\includegraphics[width=7.5cm]{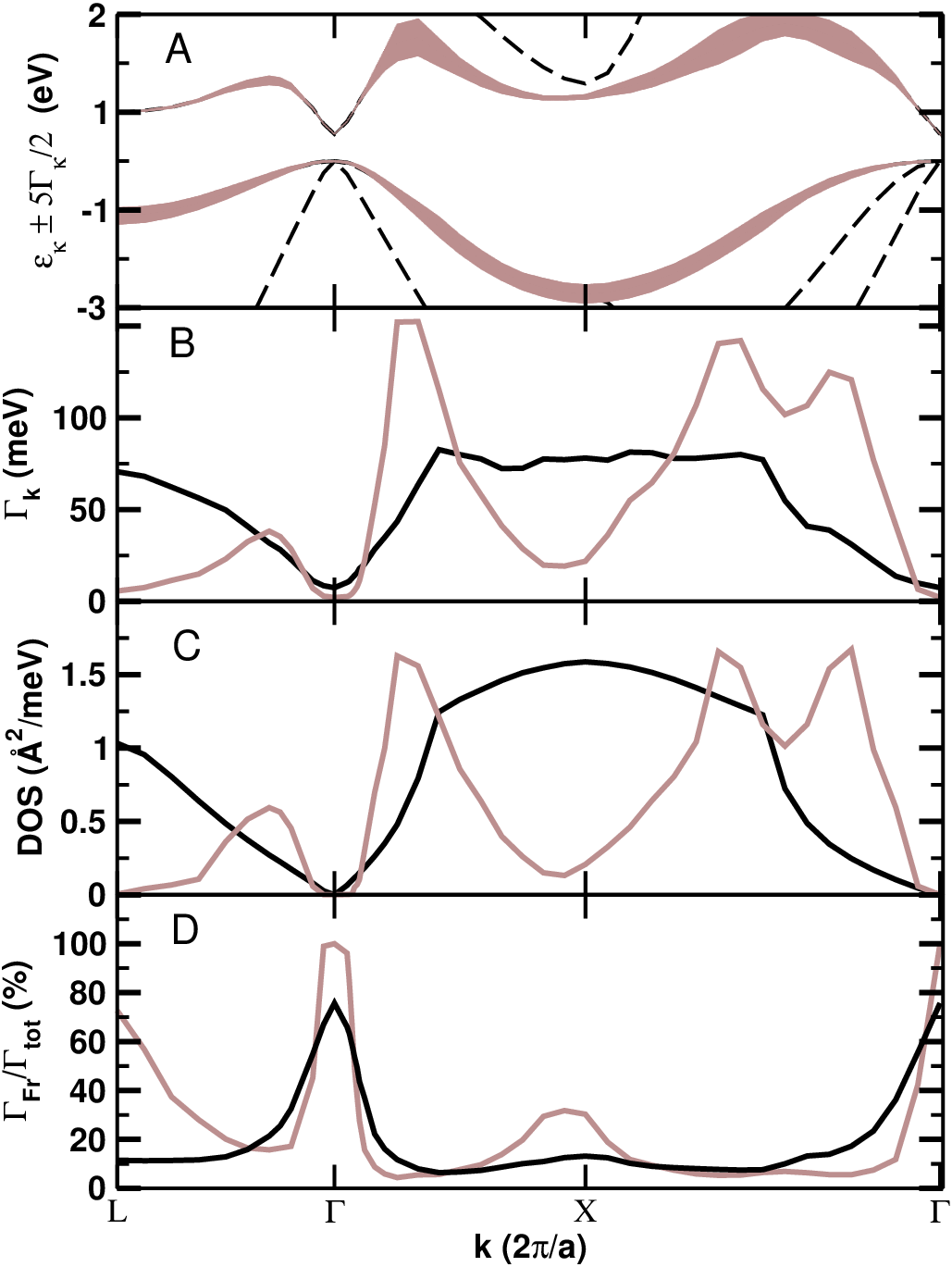}
\caption{GaAs. Panel A: highest valence and lowest conduction bands broadened by  the electron-phonon interaction.
The broadening is due to all electron-phonon interaction processes. 
For the sake of visibility, the calculated broadening $\Gamma_{\mathbf{k}}$ of the electronic bands has been multiplied by 5 in this panel. 
For panels B-D, the black solid line describes the behaviour of the initial electronic state in the highest valence band and the grey solid line is 
that of the initial state in 
the lowest conduction band.
Panel B: broadening $\Gamma_{\mathbf{k}}$ due to electron-phonon interaction, as a function of the $\mathbf{k}$ vector of the
initial electronic state. 
Panel C: electronic density of the final states allowed by conservation laws, as a function of the $\mathbf{k}$ vector of the 
initial electronic state.  
Panel D: relative contribution of the Fr\"ohlich interaction only to the total broadening.  
The calculations were done at $T=$300~K.
} 
\label{fig:width} 
\end{center}
\end{figure}

The method of interpolation of the electron-phonon matrix elements in the Wannier space is necessary when one has to calculate
integrals involving many $\mathbf{q}$ points, as it allows to significantly reduce the computational cost, compared to
direct DFPT calculation. 
We have applied the method described in previous section to calculate the total probabilities of the electron-phonon scattering
for an electron initially in the lowest conduction band of GaAs, and for a hole initially in the highest valence band of GaAs. 

In figure~\ref{fig:width}, we show the full width at half maximum $\Gamma$ due to the electron-phonon coupling, which was calculated for the lowest conduction band and the
highest valence band of GaAs as a function of the $\mathbf{k}$ vector of the
initial electronic state, at a temperature of 300~K: 
\begin{eqnarray}
\Gamma_{n\mathbf{k}}=\frac{2\pi}{\hbar}\sum_{n'}\sum_{\nu}\int_{BZ}d\mathbf{q}|g_{nn'}^{\nu}(\mathbf{k},\mathbf{k}\pm\mathbf{q})|^2 \nonumber\\
\times\delta(\varepsilon_{n'\mathbf{k}\pm \mathbf{q}}-\varepsilon_{n\mathbf{k}}\mp\hbar\omega_{\mathbf{q}\nu})
{N_{\mathbf{q}}\brace N_{\mathbf{q}}+1}
.
\label{Width}
\end{eqnarray}

Here, $\varepsilon_{n\mathbf{k}}$ are the electronic eigenenergies,
and $N_{\mathbf{q}}$ is the phonon occupation number which is described by the Bose-Einstein distribution function.
Upper and lower symbols refer to absorbtion and emission, respectively.

In practice, the delta function in eq. (\ref{Width}) was replaced by a Gaussian function in order to calculate numerically the integral in eq.~(\ref{Width}).
The integration was performed on a $48\times48\times48$ $\mathbf{q}$-point grid in the BZ.
The calculation was converged with respect to the Gaussian broadening starting from  15 meV broadening.
As the pseudopotentials used here reproduce well the respective positions of the minima of the conduction band of GaAs \cite{Sjakste:2007},
the Kohn-Sham band structure values at equillibrium were used to calculate the integral (\ref{Width}).

As one can see, the broadening due to electron-phonon coupling varies from a few meV at the bottom of the conduction band 
or at the top of the valence band (i.e. for $\mathbf{k}$ close to $\Gamma$ point of the BZ), to several tenths of meV at high initial electron energies for the conduction states, or low initial hole energies for the valence band.
The behaviour of the total electron-phonon scattering probability is similar to the one of the density of the final electronic states allowed by the energy and momentum conservation laws (panel C),
as the probability grows when more final states are available for the electron-phonon scattering. It is, however, not exactly the same, as the electron-phonon matrix elements are not constant
over the Brillouin zone.

The contribution due to the Fr\"ohlich interaction is of a few meV and does not change much over the Brillouin zone. It is, however, the dominant scattering process for the
electron close to the bottom of the $\Gamma$ or $L$ valleys, and for the hole close to the top of the valence band, as one can see from the Panel D of Fig~\ref{fig:width}. 
Away from the band extrema, the intervalley electron-phonon scattering
mechanism rapidly becomes the dominant scattering mechanism.
This result is in agreement with available literature for low-field transport \cite{Stillman:1970}. 
 Indeed,  at ambient temperatures, the Fr\"ohlich interaction is the dominant scattering mechanism which determines the low-field transport in GaAs~\cite{Stillman:1970}.
At high fields, however, the  Fr\"ohlich interaction  no longer plays the main role, and the intervalley scattering is expected to determine 
the relaxation dynamics of electrons and/or holes, as can be deduced from Fig.~\ref{fig:width}.  In this respect, GaAs  behaviour in similar to that of 
non-polar semiconductors, \textit{i.e.} silicon or germanium \cite{Jacoboni:1983}. 

\subsection{Relaxation times related to electron-phonon coupling}

The widths of the electronic levels due to electron-phonon coupling presented on  Fig.~\ref{fig:width}
can be used to estimate the relaxation times of hot electrons related to electron-phonon scattering.
 Recently, relaxation time of hot electrons excited in the CB of GaAs close to $\Gamma$ at excess energy 
$\epsilon_{ex}=0.78$ eV with respect to 
the CB bottom was found to be 22$\pm$3 fs at 293 K \cite{Kanasaki:2014}.  
We find the electron-phonon scattering time 
$\tau_{\mathbf{k}}=\frac{\hbar}{\Gamma_{\mathbf{k}}}$ to be 30 fs at $\epsilon_{ex}=0.78$ eV, 
in satisfactory agreement with the experimental result of Ref.~\onlinecite{Kanasaki:2014}. 

\subsection{Broadenings of some critical points}

\begin{figure}
\begin{center}
\includegraphics[width=7.5cm]{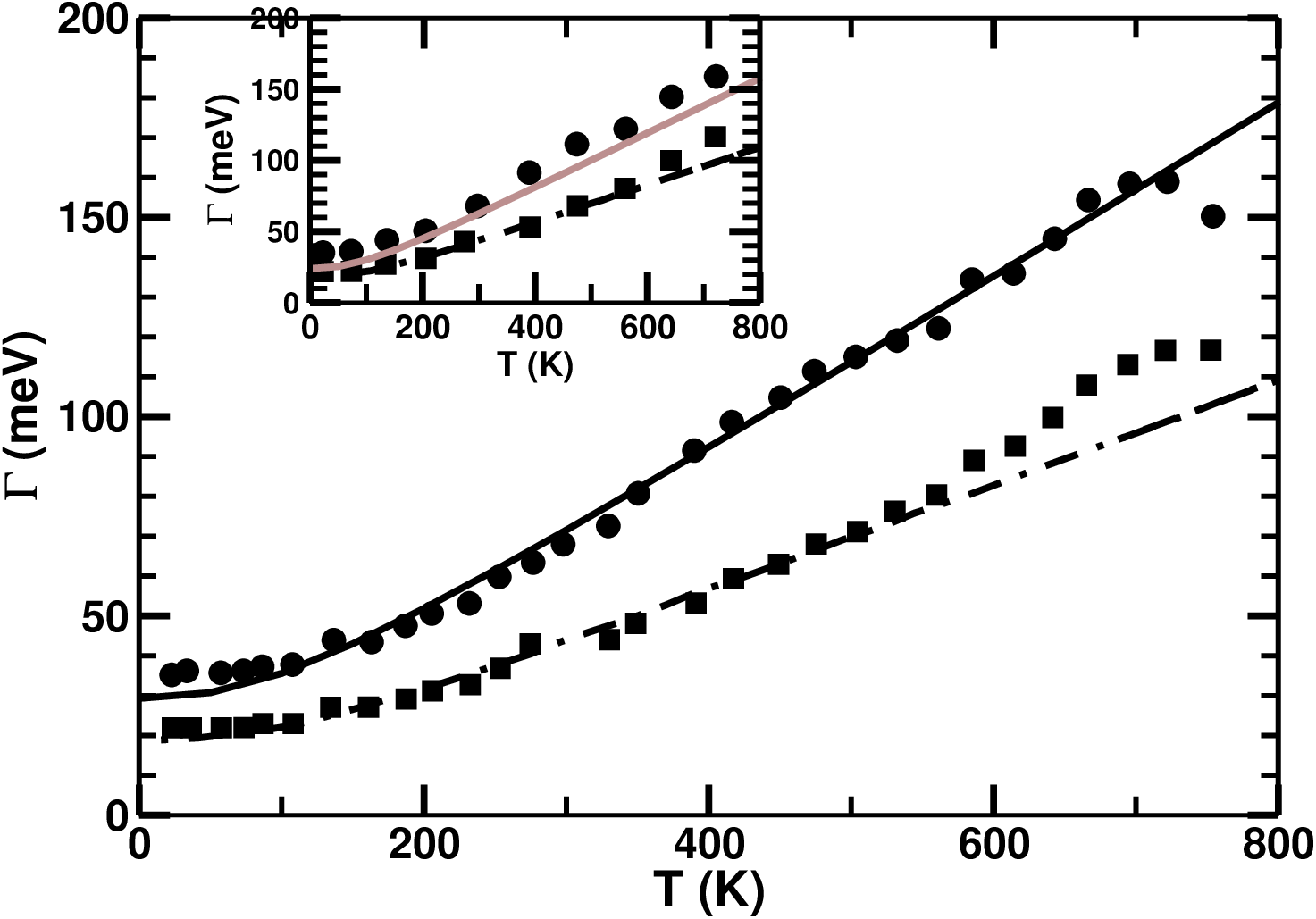}
\caption{GaAs. Broadening (meV) of the  critical point $E_1$ as a function of temperature. 
Points: experimental data of Ref.~\protect\onlinecite{Gopalan:1987}
obtained by fit with
the Fano-type excitonic line \cite{Lautenschlager:1987}. Squares: experimental data of Ref.~\protect\onlinecite{Gopalan:1987}, fit with
the two-dimensional critical point model\cite{Lautenschlager:1987}. 
Solid black line: this work, Wannier interpolation method extended to polar-optical coupling. 
Dashed-dotted line: theoretical calculation from Ref.~\protect\onlinecite{Gopalan:1987} with empirical pseudopotential method.
Both theoretical broadenings are due to electron-phonon coupling only and are    
calculated as the sum of the broadenings of the lowest conduction and highest valence bands, averaged over the 
four points along the $\Lambda$ direction as explained in the text. 
The inset represents the same experimental and theoretical results of Ref.~\protect\onlinecite{Gopalan:1987} as the
main figure, with the same notations. Grey solid line on the inset figure: this work,  Wannier interpolation  
method without extension to polar-optical coupling.
}
\label{fig:critical:E1}
\end{center}
\end{figure}

\begin{figure}
\begin{center}
\includegraphics[width=7.5cm]{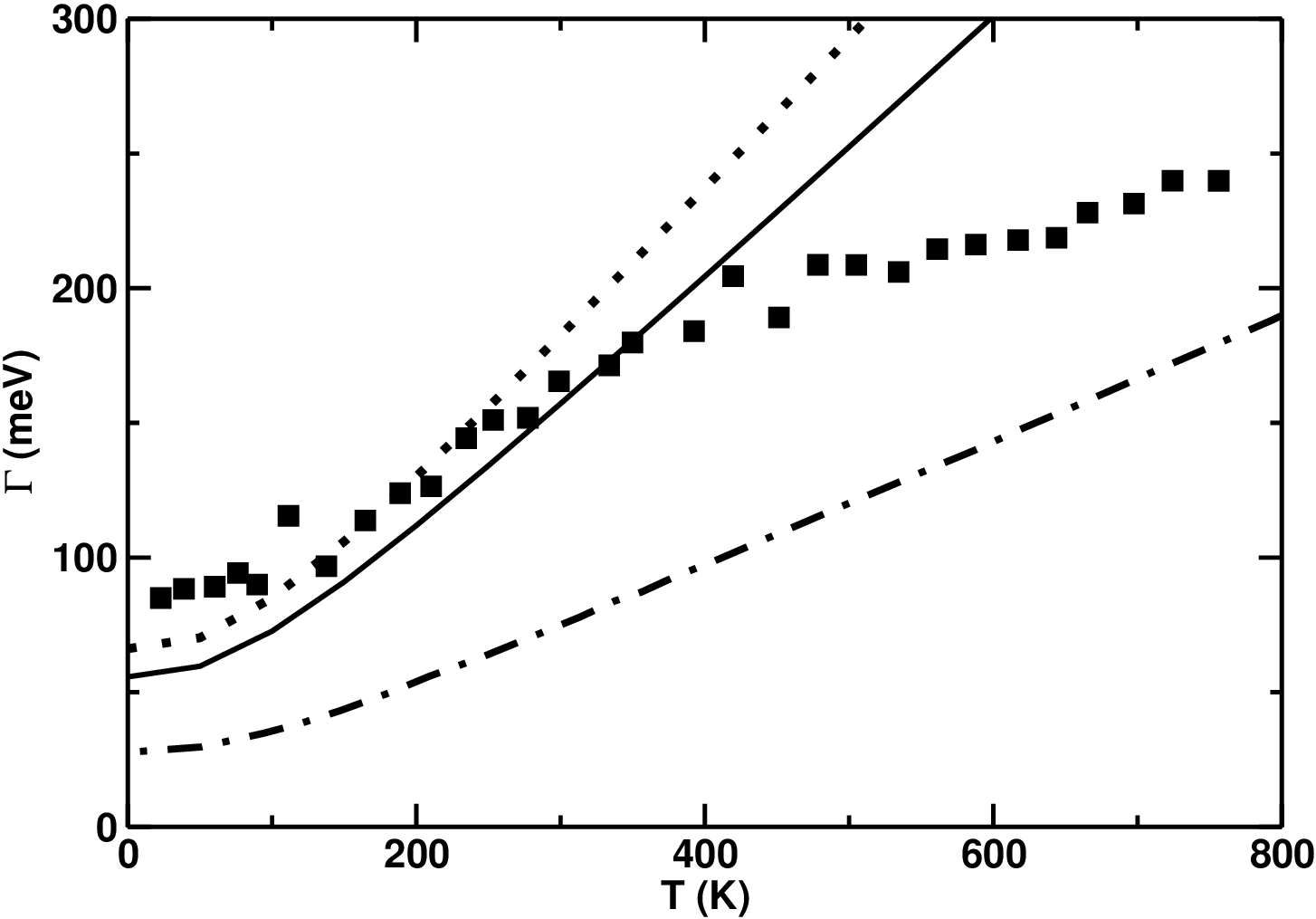}
\caption{GaAs. Broadening (meV) of the  critical point $E_2$ as a function of temperature.
Squares: experimental data of Ref.~\protect\onlinecite{Gopalan:1987}, fit with
two-dimensional critical point model\cite{Lautenschlager:1987}. 
Dashed-dotted line: theoretical calculation from Ref.~\protect\onlinecite{Gopalan:1987} with empirical pseudopotential method.
Solid black line: this work, Wannier interpolation method extended to polar-optical coupling, 
sum of the broadenings of the lowest conduction and highest valence bands, averaged over three points (see text).}
\label{critical:E1}
\end{center}
\end{figure}

The widths of the conduction and valence bands  
presented in Fig.~\ref{fig:width} can be also used to estimate the temperature-dependent 
broadenings of excitons at critical points in GaAs, attributed mostly to electron-phonon scattering. 
In principle, one should rely on the many-body excitonic wavefunction to obtain the excitonic lifetime. 
The latter consists in a combination of products of  electron and hole quasiparticle (QP) wavefunctions $\Psi_{e}  \Psi_{h}$,  
linearly mixed through the exchange operator - the electron-hole interaction \cite{Onida:2002}. 
Obtaining a precise knowledge of the QP wavefunctions and of the coefficients of the linear combination is out of the scope of this work. 
Instead,  
following Ref.~\onlinecite{Gopalan:1987}, we model the excitonic wave function
 using the  DFT wavefunctions of the lowest conduction band for the electron (resp. highest valence band for the hole) and take 
 $\Sigma_{\mathbf{k}=\mathbf{k}_i}\psi_{\mathbf{k}c}  \psi_{\mathbf{k}v}$ as our excitonic wavefunctions, with the sum 
limited to a few representative points $\mathbf{k}_i$.  
We then approximate the excitonic lifetime by the sum 
of the electron and hole lifetime, $\Gamma_{\mathbf{k}c} +  \Gamma_{\mathbf{k}v}$. 
For the  critical point $E_1$, 
the width  $\Gamma_{\mathbf{k}c}$ (resp. $\Gamma_{\mathbf{k}v}$) of the electronic (resp. hole) level  are 
obtained \textit{via} eq.~(\ref{Width})  with four points $\mathbf{k}_i$ equal to $L$,$\frac{3}{4}L$,$\frac{1}{2}L$ and
$\frac{1}{4} L$,  as done in Ref.~\onlinecite{Gopalan:1987}. 
In the case of $E_2$, only one representative point $\mathbf{k}=2\pi/a(\frac{3}{4},\frac{1}{4},\frac{1}{4})$
was used in Ref.~\onlinecite{Gopalan:1987}. The region in the $\mathbf{k}$ space where valence and conduction bands are parallel and
which contributes to the $E_2$
point was described in Ref. \onlinecite{Alouani:1988}. In this work, we decided to take into account three points 
$\mathbf{k}_i$: $(\frac{3}{4},\frac{1}{4},\frac{1}{4})$, $(1,\frac{1}{8},\frac{1}{8})$ and $U$ which belong to the
region which contributes to $E_2$\cite{Alouani:1988}. 

\subsubsection{Broadening of the $E_1$ critical point}
The resulting broadening $\Gamma_{\mathbf{k}c} +  \Gamma_{\mathbf{k}v}$, is reported as a function of temperature for the critical point $E_1$ (Fig.~\ref{fig:critical:E1}). 
As one can see,  
broadenings of the  critical point $E_1$ calculated in this work are in satisfactory agreement with the experimental results of Ref.~\onlinecite{Gopalan:1987}.
In the case of our calculation, the agreement is best with the experimental data obtained by fit with the Fano-type excitonic shape, whereas the
previous theoretical result of  Ref.~\onlinecite{Gopalan:1987} privileged the fit with two-dimensional critical point model\cite{Lautenschlager:1987}.  
The question to discriminate between the two methods of fit of the experimental data is, however, out of the scope of present work. Indeed, here we only demonstrate
that the calculated widths due to electron-phonon scattering allow to correctly estimate the magnitude of the broadenings of critical points, within the experimental 
error bar.

In the inset of Fig. \ref{fig:critical:E1}, we show the result of the calculation of the same broadening of $E_1$, but with the standard interpolation method. 
As one can see, the broadening of $E_1$ is slightly lower if the Fr\"ohlich coupling is omitted, however, the overall result is very similar, confirming 
the statement of the previous section that the scattering of the electrons/holes away from valley minima is dominated not by the Fr\"ohlich, but by the 
intervalley scattering. 
 
\subsubsection{Broadening of the $E_2$ critical point}
In Fig.~\ref{critical:E1}, we have estimated the broadening of the  critical point $E_2$ in GaAs.
In the case of $E_2$, our results (solid line) are very different from the theoretical result obtained with the empirical pseudopotential method\cite{Gopalan:1987},
and yield a much better agreement with experiment, showing that the experimentally measured broadening of the $E_2$ point can be attributed 
to the electron-phonon scattering. It must be noted down, however, first that only  experimental results  extracted with the fit with the two-dimensional critical
point model are
available in this case. Second, the experimental behaviour of the broadening beyond 500 K differs from the one predicted by the calculation.  

In dotted line, we show the broadening of the  critical point  $E_2$ estimated in our work with only one 
point $\mathbf{k}=2\pi/a(\frac{3}{4},\frac{1}{4},\frac{1}{4})$ as was done in Ref.~\onlinecite{Gopalan:1987}. As one can see,
the broadening reported  with dotted line is similar to the one obtained with three representative points, showing that
the result does not depend crucially on the method of averaging and that it remains widely different from the
one obtained with empirical pseudopotential.

\section{Conclusion}

In conclusion, we have presented the description of the extension to polar-optical coupling of the
method which allows to interpolate the electron-phonon matrix elements in the space of maximally-localized Wannier functions.
The extended method is based on Vogl's model of the Fr\"ohlich component of the electron-phonon coupling, and allows to
interpolate the electron-phonon matrix elements calculated within DFPT in polar semiconductors with excellent precision.
We have applied the extended method of interpolation in the case of GaAs, and calculated the widths of the electronic 
levels due to the electron-phonon coupling for highest conduction and lowest valence band. We have demonstrated that 
the obtained widths of the electronic
levels can be used to estimate the relaxation times of hot electrons and the broadenings of the critical points due
to electron-phonon scattering, in good agreement with various experiments. Finally, we have shown that, although the Fr\"ohlich
interaction is the dominant scattering process for electrons/holes close to the valley minima, in agreement with low-field 
transport results, at higher energies, the intervalley scattering is expected to dominate the relaxation dynamics of hot electrons 
or holes.

\section{Acknowledgments}
This work was supported by the French ANR (project PNANO ACCATTONE)
and by the Graphene Flagship. 
Results have been obtained with  \textsc{Quantum ESPRESSO} package~\cite{QE-2009,Baroni:2001} and Wannier90 package \cite{Marzari:2012}.
We acknowledge support from the French DGA, and computer time has been granted by GENCI (project 2210) and by Ecole Polytechnique through the LLR-LSI project. 
J. Sjakste and N. Vast acknowledge with gratitude many  discussions with Prof. V. Tyuterev, from Tomsk Pedagogical University, Russia,  
on the possibility to interpolate electron-phonon matrix elements in polar materials.

\end{document}